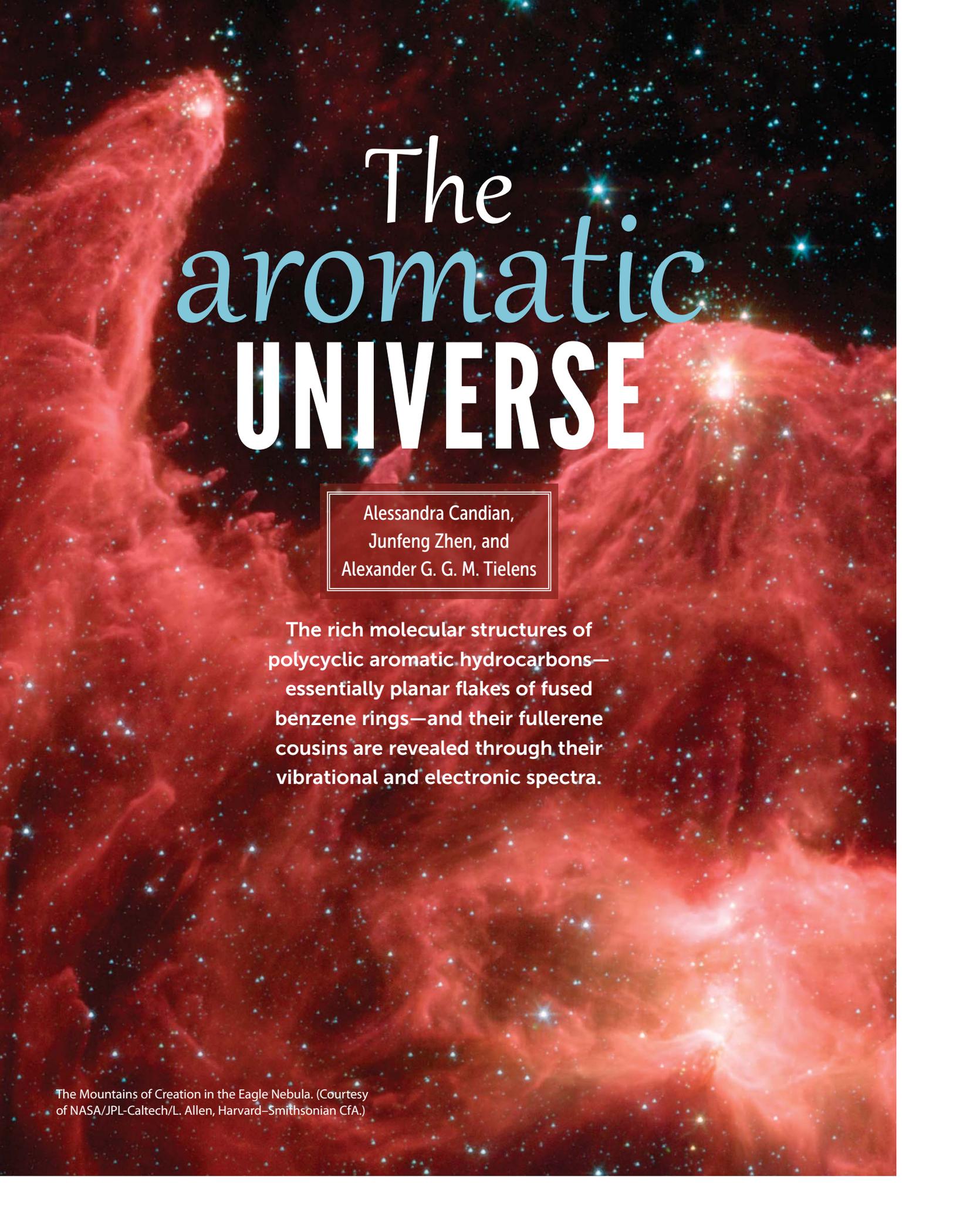

# The *aromatic* UNIVERSE

Alessandra Candian,
Junfeng Zhen, and
Alexander G. G. M. Tielens

The rich molecular structures of polycyclic aromatic hydrocarbons— essentially planar flakes of fused benzene rings—and their fullerene cousins are revealed through their vibrational and electronic spectra.

The Mountains of Creation in the Eagle Nebula. (Courtesy of NASA/JPL-Caltech/L. Allen, Harvard–Smithsonian CfA.)

**Alessandra Candian** is a Veni Research Fellow and **Xander Tielens** is a professor of physics and chemistry of the interstellar medium, both at the Leiden Observatory at Leiden University in the Netherlands. **Junfeng Zhen** is an assistant professor of astrochemistry in the department of astronomy at the University of Science and Technology of China in Hefei.

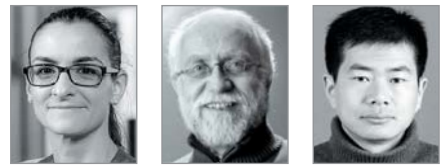

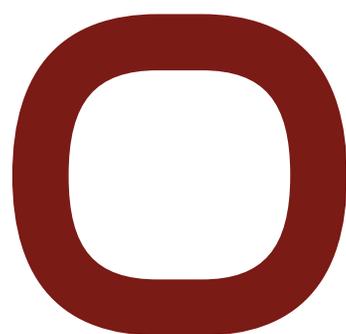

Over the past 20 years, ground- and space-based observations have revealed that the universe is filled with molecules. Astronomers have identified nearly 200 types of molecules in the interstellar medium (ISM) of our galaxy and in the atmospheres of planets; for the full list, see www.astrochymist.org. Molecules are abundant and pervasive, and they control the temperature of interstellar gas (see the box on page 40).[1] Not surprisingly, they directly influence such key macroscopic processes as star formation and the evolution of galaxies.

Interstellar molecules attest to the chemical evolution of the universe—from the atoms formed in stellar interiors to the dust and planets of solar systems. Besides controlling the birth and evolution of stars and galaxies, the molecules can be used as unique probes of conditions in the ISM and as tracers of processes therein. Among those molecules are polycyclic aromatic hydrocarbons (PAHs)—essentially fused benzene rings joined together in flat sheets whose edges are decorated with hydrogen atoms. The multiple benzene-ring configuration delocalizes the electrons in the rings and stabilizes the PAHs.

On Earth, PAHs are common by-products of the incomplete combustion of organic material, such as coal and tar. Among other manifestations, they appear as part of the burned crust on your barbecued meat, and their stability makes them a major pollutant of air and water.

In space, the molecules' spectral fingerprints are observed seemingly everywhere, from dusty disks around young stars and regions near massive stars to the ISMs of galaxies, and they have been observed as far back in time as when the universe was a mere 3.3 billion years old. Those fingerprints are produced when PAH molecules become highly excited after absorbing UV photons emitted by the stars and then cool through vibrational emission of IR photons. Estimates of the strength of that emission indicate that PAHs contain up to 15% of all the carbon in the ISM.

The recent finding of the buckminsterfullerene molecule ($C_{60}$) in space[2] adds complexity to the inventory of known interstellar carbonaceous molecules beyond PAHs—chains and nanodiamonds among other allotropes. The presence of $C_{60}$ in regions where PAHs have also been detected suggests that PAHs and fullerenes may be related by similar synthetic mechanisms in space. The similarity heralds some fascinating chemistry that scientists have only started exploring.

In this article we discuss the status of the aromatic universe. It's an exciting time: IR astronomy will get a huge boost when NASA's *James Webb Space Telescope* (*JWST*) is put into Earth's orbit in 2021. And the Extremely Large Telescope, which can resolve wavelengths from the UV to the mid-IR, is currently under construction in Chile. Let the journey begin.

## PAHs in space

The mid-IR spectra of regions associated with UV-illuminated dust and gas show a series of strong emission features, generally known as the aromatic infrared bands (AIBs). Those bands, shown in figure 1, are the signatures of PAH molecules near the gas clouds in the Orion Nebula about 1200 light-years from Earth. It is widely accepted among astronomers that far-UV photons excite large PAHs—each typically with 50–100 C atoms. Upon excitation, the molecules quickly relax by IR fluorescence through their vibrational degrees of freedom.

The main features, observed at 3.3, 6.2, 7.7, 8.6, 11.2, 12.7, and 16.4 μm, correspond to the vibrational modes of aromatic molecules. As labeled at the top of the spectra in the figure, those modes primarily include the stretching of C–H and C–C bonds, combinations of those stretches, and in-plane and out-of-plane C–H bending motions. The spectroscopic identification of secondary bands at 3.4, 5.7, 11.0, 13.5, and 17.4 μm is still open to question. All the bands are perched on broad plateaus attributed to clusters of PAHs.[1]

PAH molecules responsible for the AIBs are quite sensitive to the local environment's physical conditions, such as the intensity of the radiation field. Indeed, the AIBs show spectral variations not only among different astronomical objects,[3] such as dense clouds of molecular gas or planetary nebulae—the outermost shell of ionized gas ejected from old red giant stars late in their lives—but also in the same object.[4] Those variations affect not only the relative intensities of the different bands but also their peak positions and profiles.

Astronomers have classified the profile variations of a large number of objects observed by the *Spitzer Space Telescope* into four classes, A–D.[5,6] The different classes are associated with particular source types: Class A sources are interstellar matter illuminated by a star—the general ISM of a galaxy falls into that category; class B sources consist of stars illuminating their own circumstellar material, as in the case of a young star illuminating its protoplanetary disk; and class C and D sources are typically C-rich older stars.

PAH molecules belong to a large and diverse chemical family. To understand them, researchers analyze the characteristic spectral features of the PAHs. In fact, even if the observed vibrational modes are typical of a family of molecules rather than a specific molecule, some features, such as electric charge or the presence of functional, or chemical, groups, leave a telltale signature in the vibrational spectrum.

Our knowledge of the PAH populations in space comes from the concerted efforts of astronomers, theoretical chemists, and experimentalists who have worked together to create tools, such as the NASA Ames PAH database (www.astrochemistry.org/pahdb) and the French–Italian PAH database (http://astrochemistry.oa-cagliari.inaf.it/database), that describe hundreds of PAHs. We now know that those astronomical PAHs, known as astroPAHs, are overwhelmingly aromatic—that is, the edges of the fused benzene rings are decorated almost completely by H atoms. Depending on the intensity of the illuminating light and the density of the medium, astroPAHs can exist in a range of charged states—from nega-tive to doubly positive—and may have either lost a few peripheral H atoms or gained new ones.[6] They may also contain traces of some other atomic species, such as nitrogen, that can replace C atoms in the ring.

Although the astroPAH family appears to be quite diverse, circumstantial evidence points to a small group of molecules, the grandPAHs, that dominate the population. Large and hence quite stable molecules that can survive the harsh conditions of the ISM, they are thought to be the reason for the almost identical emission spectra seen in most interstellar sources and in the simplicity of the AIB spectrum around 15–20 μm. The evolution of PAH features near the central

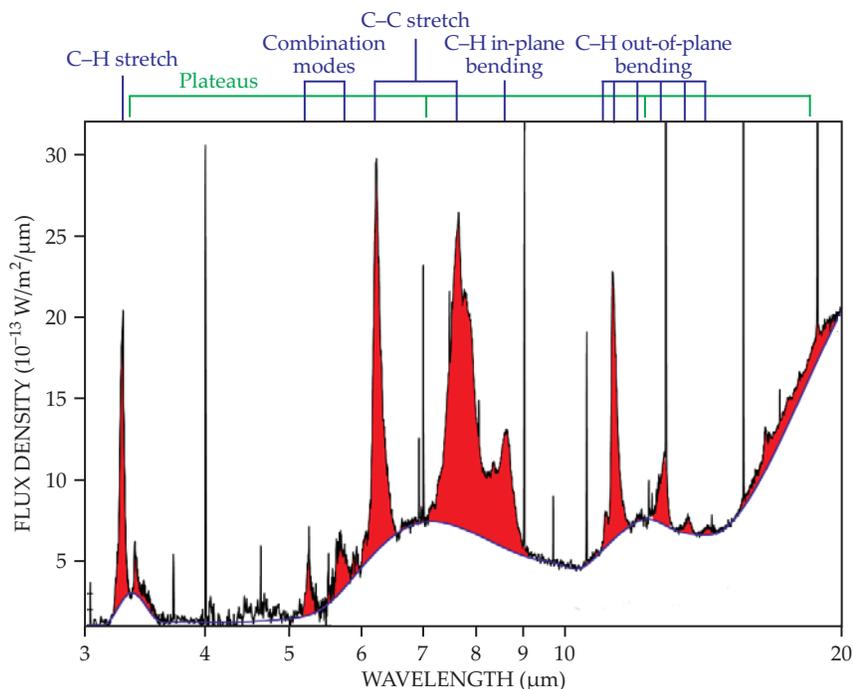

FIGURE 1. THE MID-IR SPECTRA of the famous photo-dissociation region of the Orion Nebula as observed by the European Space Agency's *Infrared Space Observatory*. The red emission peaks correspond to the vibrational modes of polycyclic aromatic hydrocarbon (PAH) molecules. The specific bond stretches associated with those modes are labeled at the top, and they are perched on broad plateaus attributable to clusters of PAHs. (Adapted from ref. 3, Peeters et al.)

## THE INTERSTELLAR MEDIUM

The space between stars in a galaxy—a region known as the interstellar medium (ISM)—is filled with gas, dust, cosmic rays, and electromagnetic radiation. Its matter comes in different phases that depend on the temperature, density, and composition of the ions, atoms, and molecules found there. The temperature and density, in turn, are governed by other factors. One of the most important mecha-nisms to heat the gas is the photoelectric effect, by which far-UV photons are absorbed by dust grains or large molecules that release electrons with excess kinetic energy in the gas phase.

Yet a warm gas can also shed energy via the different energy transitions available to its molecules. Indeed, clouds in the ISM that contain predominantly molecular gas are among the coldest places in the universe.

The ISM is an essential element of a galaxy because it acts as an intermediary between the stars and the galaxy that contains them. Stars form in the densest region of the ISM, and when they die they replenish the ISM with matter that fuels the next generation of stars. That interaction determines the rate at which a galaxy can form stars. (See the article by Christoph Federrath, PHYSICS TODAY, June 2018, page 38.)

young star in the Iris Nebula gives credibility to that hypothesis, although efforts to fit the typical spectrum of an interstellar source with only a few highly stable PAH molecules have not yet been successful.[7]

## Fullerenes

In 1985 Robert Curl Jr, Harold Kroto, and Richard Smalley discovered $C_{60}$ in the molecular clusters produced by laser vaporization of graphite in Smalley's lab at Rice University. The work earned the three chemists a Nobel Prize (see PHYSICS TODAY, December 1996, page 19), but it was a serendipitous find. The inspiration for the experiment came from Kroto's interest in proving that long-chain C molecules are born in the stellar atmospheres in red giants. Concurrent with the laser vaporization experiments, astrochemists were working with spectroscopists to study IR emissions from those stars. Unsurprisingly, as soon as $C_{60}$ was discovered in graphite vaporization, its presence in space was immediately postulated. But actual detection remained elusive for 25 years.

In 2010 astronomer Jan Cami at the University of Western Ontario and his colleagues took an IR spectrum, shown in figure 2a, of the young planetary nebula Tc 1. Instead of the PAHs usually observed in such nebulae they found strong vibrational transitions at 7.0, 8.6, 17.4, and 18.9 μm. The researchers recognized the peaks as a close match for $C_{60}$ and $C_{70}$ and realized that the H-poor conditions in Tc 1 could be behind the production of the fullerenes;[2] later studies, however, showed that Tc 1 is not particularly H poor.

Since 2010, fullerene molecules have usually been identified by that 18.9 μm peak in planetary nebulae, in star-forming regions, in young stellar objects, and in the diffuse ISM—often together with PAH emissions.[8] What's more, they represent a significant (roughly 0.001%) source of C in the universe.

In 2015 at the University of Basel, John Maier added another chapter to the story of interstellar fullerenes when he, Ewen Campbell, and their collaborators[2] unequivocally assigned four diffuse interstellar bands (DIBs) to the near-IR electronic transition of the ion $C_{60}^+$. Shown in figure 2b, the DIBs are a family of around 500 mostly unidentified absorption features—from the near-UV to the near-IR—in the spectra of interstellar clouds in the Milky Way and other galaxies. It had been widely accepted that DIBs came from electronic transitions in molecules, but their identification had completely eluded astronomers until Maier's work. Puzzles abounded: For one thing, although those bands roughly correlate with each other in strength, there are weak variations from star to star. To identify $C_{60}^+$ as the "carrier" of the absorption bands, it became necessary to combine laboratory studies, theoretical modeling, and astronomical observations of the DIBs.

Maier and colleagues' experiment on $C_{60}^+$ offered a close spectral match to the DIBs' electronic transitions at 9577 Å and 9632 Å, when the lab spectra were measured under astrophysically relevant conditions. Whereas vibrational spectroscopy cannot perfectly distinguish between molecules of similar chemical bonds, electronic spectroscopy can provide fingerprints of individual molecules. Other fullerenes besides $C_{60}$ and $C_{70}$ may also reside in space, but researchers need to further investigate the individual cases.

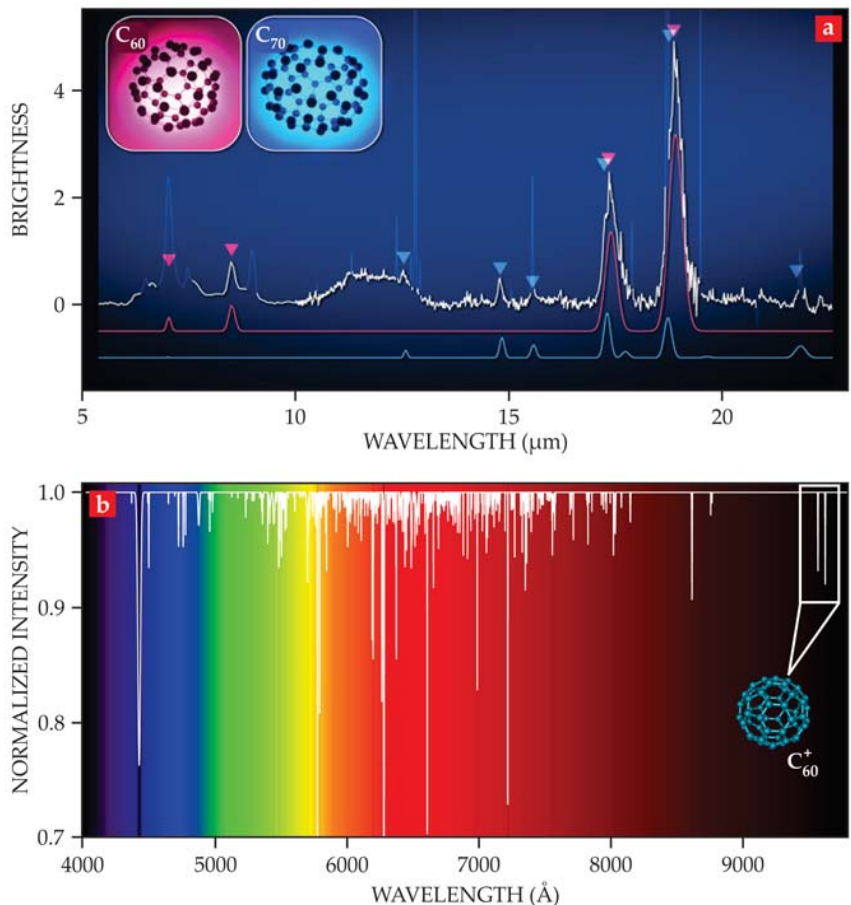

**FIGURE 2. THE FINGERPRINTS OF BUCKYBALLS. (a)** Carbon-60 (pink) and its elongated relative $C_{70}$ (blue) appear in the IR vibrational spectrum of the young planetary nebula Tc 1. The spectrum, taken by NASA's *Spitzer Space Telescope*, represents the first evidence that fullerenes can form in interstellar space. (Adapted from ref. 2, Cami et al.) **(b)** This synthetic absorption spectrum of the diffuse interstellar medium illustrates the richness of the so-called diffuse interstellar bands (DIBs) across the optical and near-IR regions of the electromagnetic spectrum. In 2015 Ewen Campbell and collaborators proved that the singly charged $C_{60}^+$ ion is responsible for two strong (9577 Å and 9632 Å) and two faint (not shown) DIBs of the spectrum.[2] (Image courtesy of Jan Cami.)

## Bottom-up chemistry

Gas-phase chemistry in space is usually thought to involve the buildup of molecules a few atoms at a time in a so-called bottom-up process. For PAHs, the limiting step in that process is the formation of the first aromatic ring, the benzene molecule $C_6H_6$. But astrochemists have proposed few routes to account for its presence in the ISM.

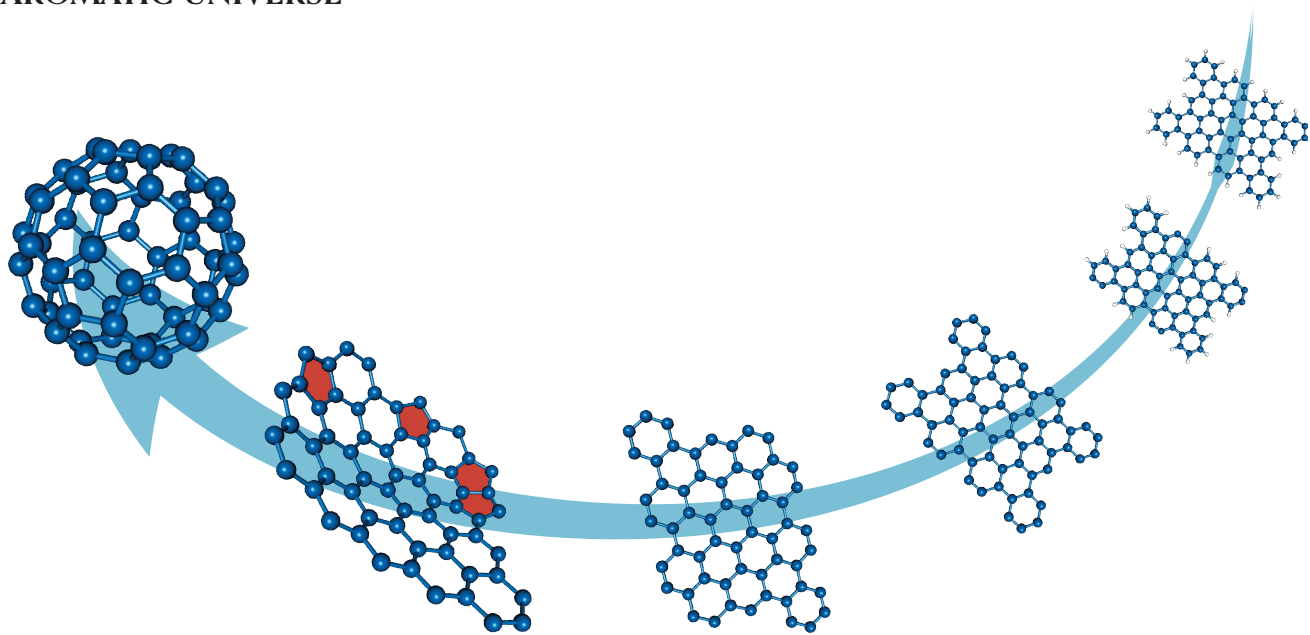

**FIGURE 3. TOP-DOWN CHEMISTRY IN SPACE.** In this artistic impression, large polycyclic aromatic hydrocarbon (PAH) molecules exposed to the strong radiation field of a star first lose all their peripheral hydrogen atoms (white atoms, top right) and are transformed into small graphene flakes whose fragile, dangling carbon rings at the corners break off. That degradation is then followed by the loss of carbon atoms. The loss creates pentagons (red) in the dehydrogenated PAH molecule, which bends the structure out of the plane and culminates in the formation of a $C_{60}$ fullerene.

Benzene could be produced from the closing up of C chain molecules, some suggest.[9] But because of the low (roughly 10 K) temperatures of the ISM, the reactions required to form benzene and larger PAHs would need to be barrierless. In 2011 Brant Jones of the University of Hawaii at Manoa and collaborators discovered such a barrierless reaction for converting $C_2H$ and $H_2CCHCHCH_2$ to benzene.[10] Subsequent additions of vinylacetylene ($CH_2CHCCH$) to that first ring can then trigger the growth of PAHs.[11] Benzene can also react with CN to create the aromatic benzonitrile, which was detected in tiny amounts in the ISM earlier this year.[12]

Although those gas-phase, bottom-up reactions are interesting possibilities, some issues remain unresolved. For example, the vinylacetylene addition requires that the benzene ring lose an H atom, a process unlikely to happen in the shielded, low-UV environments where benzonitrile has been detected. Moreover, the vinylacetylene molecule itself is not expected to be abundant in the same regions. And given the low abundance of that rather small species, it would be difficult to build PAH molecules containing 50 C atoms or more.

The abundance of large PAHs and fullerenes in space therefore calls for a completely different synthetic chemistry. Interstellar PAHs are generally thought to form in the stellar ejecta of so-called asymptotic giant branch stars, whose density and temperature lead to chemistry akin to what takes place in a combustion engine.[13] Those stars eject much of their C into space mostly in the form of soot. Interstellar PAHs are then either the molecular intermediaries in or by-products of the soot formation process. Those molecular species are carried by the gentle stellar winds and mixed into the general ISM. As PAHs are then further exposed to the ISM, energetic photons and cosmic rays weed out the less stable species and convert others into stable ones.

In recent years, observational evidence has emerged in support of that scenario and, more specifically, the intertwining of PAH and fullerene chemistry in space. (For a discussion of chain reactions that may explain soot inception and growth, see the news story on page 18.) IR studies using the *Spitzer Space Telescope* and the *Herschel Space Observatory* have shown that as matter approaches a UV-bright star, the abundance of PAHs slowly decreases; simultaneously, the amount of $C_{60}$ increases.[14]

## Top-down chemistry

In very harsh environments, densities are too low and time scales too short for a bottom-up mechanism to produce fullerenes. In 2012 Olivier Berné and one of us (Tielens) proposed that PAHs are highly excited by UV photons and then fragment via a top-down chemistry that creates smaller molecules from larger ones.[14] Because the C–H bond is the weakest, H atoms are the first to be stripped off. That loss is normally balanced by reactions of PAHs with nearby H atoms.

Close to the star, however, the strong radiation field may overwhelm the back reaction, and a PAH can lose all its H atoms and be transformed into a graphene-like C flake. Further UV exposure will then strip off the more strongly bound C atoms, an action that is thought to lead to the formation of pentagons. The presence of pentagons then warps the molecules out of the plane and prompts the flake to curl up into C cages and fullerenes. Figure 3 illustrates the progression of those structures.

Many details of the chemistry behind that stripping and curling are sketchy, and researchers have yet to characterize the structures of key molecular intermediaries. Even so, experiments support the general scenario of top-down interstellar chemistry.[15] The UV irradiation of PAH cations in an ion trap reveals that large PAHs are quickly stripped of all their H atoms before the PAHs start to lose C atoms, two at a

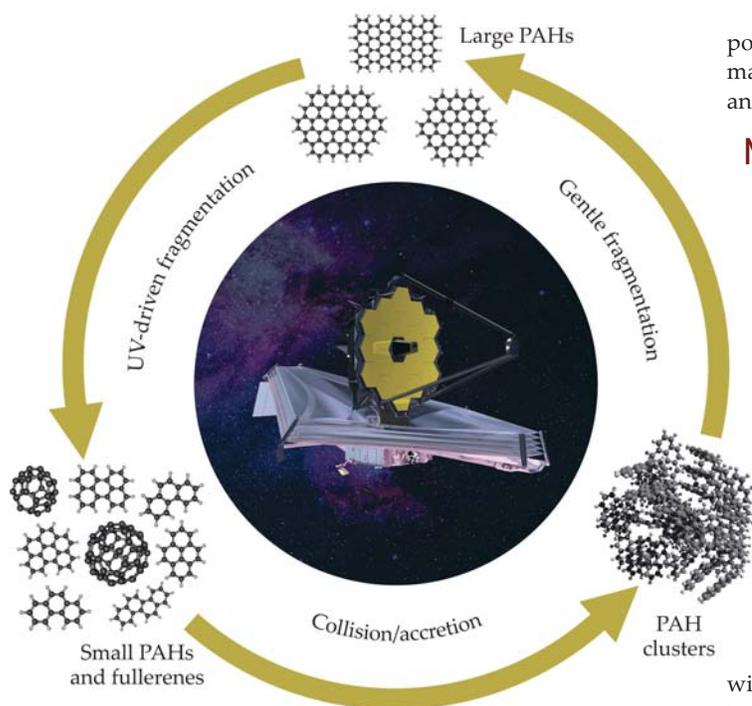

Large PAHs

Gentle fragmentation

UV-driven fragmentation

Collision/accretion

Small PAHs and fullerenes

PAH clusters

**FIGURE 4. THE LIFE CYCLE OF LARGE CARBONACEOUS MOLECULES** in the interstellar medium. Large polycyclic aromatic hydrocarbons (PAHs, top) are fragmented into small PAHs and fullerenes (bottom left) under exposure to UV photons. At the same time, those small PAHs and fullerenes can form molecular clusters (bottom right) via collisions and reactions with each other. The clusters are held together by weak van der Waals forces. Collisions or absorption of low-energy photons breaks down the clusters into PAH-like fragments that quickly react to form large PAHs again. When it is launched in 2021, the *James Webb Space Telescope* (center) will help to better characterize the life cycle.

time. The mass-spectroscopic pattern of the clusters reveals that some species are more stable, and thus more abundant, than others.

The UV processing and atom stripping occur at almost all wavelengths in large PAHs and fullerenes, with a few notable exceptions. Under laser exposure, $C_{60}$ does not absorb at 532 nm and remains perfectly stable even at very high laser intensities. Carbon-70 absorbs at that wavelength, though, and its cage eventually shrinks into the $C_{60}$ structure through the loss of C atoms before further photochemical evolution stops.[15]

The evolution of a large PAH, such as $C_{66}H_{26}$, starts with the loss of all hydrogens during UV exposure, followed by the loss of carbons. In 2010 Andrey Chuvilin and colleagues showed electron microscopy evidence that planar graphene flakes are transformed into fullerenes by the same C-loss mechanism.[16] The loss of atoms at the edges of a flake produces pentagons that cause it to curl up into bowl-like structures.

One is tempted to speculate that those top-down processes behind PAH breakdown initiated by UV irradiation or energetic particle bombardment are balanced in the interstellar medium by bottom-up growth processes, as outlined in figure 4. Inside the benign environments of molecular clouds, small PAHs can collide and cluster into larger structures.[17] When ex-posed to strong UV fields near bright stars, the larger PAHs may then fragment and change from one isomer to another, and restart the cycle.

## New perspectives

The *JWST* is currently being outfitted with instruments well suited to study the aromatic universe at high spectral and spatial resolution. Astronomers expect that the *JWST* will be able to trace in detail the cosmic history of PAHs throughout the era of peak star formation when the universe was just 4 billion to 5 billion years old. It may even allow them to trace very distant events to a time when the universe was just 1 billion years old.

What makes the *JWST* so advantageous is that it can measure the star formation rate in dusty, highly obscured regions of IR galaxies. Such studies will also provide new insights in the role of molecules in the history of the universe. PAHs are abundant in protoplanetary disks, and the limited data currently available suggest that they change chemically while being transported from the general ISM to those protoplanetary disks.

With the *JWST*, the molecular structure of the PAH family will be revealed through IR signatures. Those signatures, in turn, will yield the rich organic inventory in regions of star and planet formation—even in the habitable zone of countless solar systems never before explored.

*Alessandra Candian appreciates support through a Veni grant from the Netherlands Organisation for Scientific Research (NWO). Junfeng Zhen appreciates support from the Fundamental Research Funds for the Central Universities of China. Studies of interstellar chemistry at Leiden Observatory are supported by the NWO as part of the Dutch Astrochemistry Network, by an NWO Spinoza Prize, and by funding from a European Commission Marie Skłodowska-Curie action to the EUROPAH consortium.*

## REFERENCES

1. A. G. G. M. Tielens, *Rev. Mod. Phys.* **85**, 1021 (2013).
2. J. Cami et al., *Science* **329**, 1180 (2010); E. K. Campbell et al., *Nature* **523**, 322 (2015).
3. E. Peeters et al., *Astron. Astrophys.* **390**, 1089 (2002); B. van Diedenhoven et al., *Astrophys. J.* **611**, 928 (2004).
4. A. Candian et al., *Mon. Not. R. Astron. Soc.* **426**, 389 (2012).
5. G. C. Sloan et al., *Astrophys. J.* **791**, 28 (2014).
6. V. Le Page, T. P. Snow, V. M. Bierbaum, *Astrophys. J.* **584**, 316 (2003); H. Andrews, A. Candian, A. G. G. M. Tielens, *Astron. Astrophys.* **595**, A23 (2016).
7. H. Andrews et al., *Astrophys. J.* **807**, 99 (2015).
8. J. Bernard-Salas et al., in *Proceedings of the Life Cycle of Dust in the Universe: Observations, Theory, and Laboratory Experiments*, A. Andersen et al., eds., Sissa Medialab (2013), p. 032.
9. R. P. A. Bettens, E. Herbst, *Astrophys. J.* **478**, 585 (1997).
10. B. Jones et al., *Proc. Natl. Acad. Sci. USA* **108**, 452 (2011).
11. D. S. N. Parker et al., *Proc. Natl. Acad. Sci. USA* **109**, 53 (2012).
12. B. McGuire et al., *Science* **359**, 202 (2018).
13. M. Frenklach, E. D. Feigelson, *Astroph. J.* **341**, 372 (1989); I. Cherchneff, J. R. Barker, A. G. G. M. Tielens, *Astrophys. J.* **401**, 269 (1992).
14. O. Berné, A. G. G. M. Tielens, *Proc. Natl. Acad. Sci. USA* **109**, 401 (2012).
15. J. Zhen et al., *Astrophys. J. Lett.* **797**, L30 (2014).
16. A. Chuvilin et al., *Nat. Chem.* **2**, 450 (2010).
17. M. Rapacioli, C. Joblin, P. Boissel, *Astron. Astrophys.* **429**, 193 (2005).  PT